\documentclass[preprint2]{aastex}
\usepackage[usenames,dvips]{color}
\usepackage{graphicx,natbib}
\usepackage{hyperref}
\usepackage{amsmath}
\usepackage{multirow}
\usepackage{epsfig}
\usepackage{lineno}
\usepackage{url}

\newcommand{\altm} {\altaffilmark}
\newcommand{\psr}{PSR~J1836+5925}

\shorttitle{X-ray pulsation of PSR J1836+5925}
\shortauthors{Lin \& Hui et al.}

\begin{document}


\title{Discovery of X-ray pulsations from \emph{next Geminga} --- PSR~J1836+5925}


\author{L. C. C. Lin\altm{1}, C. Y. Hui\altm{2}, K. T. Li\altm{3}, J. Takata\altm{4}, C. P. Hu\altm{5}, A. K. H. Kong\altm{1}, D. C. C. Yen\altm{6} and Y. Chou\altm{5}}

\email{cyhui@cnu.ac.kr}

\begin{abstract}
We report the X-ray pulsation of  $\sim 173.3$ ms for the ``{\it next Geminga}", \psr, with recent {\it XMM-Newton} investigations.
The X-ray periodicity is consistent wtih the $\gamma$-ray ephemeris at the same epoch.
The X-ray folded light curve has a sinusoidal structure which is different from the double-peaked $\gamma$-ray pulse profile.
We have also analysed the X-ray phase-averaged spectra which shows the X-ray emission from \psr\ is thermal dominant.
This suggests the X-ray pulsation mainly originates from the modulated hot spot on the stellar surface. 
\end{abstract}


\altaffiltext{1}{Institute of Astronomy, National Tsing-Hua University, Hsinchu 30013, Taiwan} 
\altaffiltext{2}{Department of Astronomy and Space Science, Chungnam National University, Daejeon, South Korea} 
\altaffiltext{3}{Department of Biological Science and Technology, China Medical University, Taichung 40402, Taiwan} 
\altaffiltext{4}{Department of Physics, University of Hong Kong, Pokfulam Road, Hong Kong, PRC} 
\altaffiltext{5}{Graduate Institute of Astronomy, National Central University, Jhongli 32001, Taiwan}
\altaffiltext{6}{Department of Mathematics, Fu Jen Catholic University, New Taipei City 24205, Taiwan}   
%\altaffiltext{7}{Laboratoire de Physique et Chimie de l'Environnement et de l'Espace - Universit$\acute{e}$ d'Orl$\acute{e}$ans/ CNRS, F-45071 Orl$\acute{e}$ans Cedex 02, France}  

\section{Introduction}
The launch of {\it Fermi} Gamma-ray Space Telescope in 2008 provides the opportunity to examine $\gamma$-rays emitted 
from compact objects, and 117 $\gamma$-ray pulsars has been detected with the first three years archival data  \citep{Abdo2013}. 
Among all the pulsars detected by {\it Fermi}, 36 were directly found by the blind search in the $\gamma$-ray band and 
34 of them were identified as radio-quiet with the threshold of 30 $\mu$Jy in radio flux density at 1400 MHz.  
%(PSR~J1741-2054 \& PST~J2032+4127 were excluded by \citealt{Camilo2009}). 
%If we include the only one radio-quiet $\gamma$-ray pulsar, Geminga (i.e., PSR~J0633+17; \citealt{Bert92}), discovered before the launch of {\it Fermi} observatory and considered a stricter threshold on those without certain radio counterpart in the 2nd {\it Fermi} Large Area Telescope (LAT) catalog of $\gamma$-ray pulsars, there are still 34 $\gamma$-ray pulsars classified as radio-quiet, and PSR~J1907+0602 of the pseudo-luminosity of $L_{1400} \sim 0.035$ mJy kpc$^2$ with the assumed distance of  $3.2\pm 0.6$~kpc becomes the most radio-faint $\gamma$-ray pulsar \citep{Abdo2010b}.
The population of radio-quiet $\gamma$-ray pulsars is expanding due to the continuous accumulation of {\it Fermi} data 
and the improvement of the searching schemes (e.g., weighting $H$ test-statistic; \citealt{Kerr2011}).
This has imposed constraints on the pulsar emission geometry which suggests the radio emission and the $\gamma$-rays are not 
produced in the same region of the pulsar magnetopshere. While the radio emission is originated from the cones emerged near the stellar
surface, the pulsed $\gamma$-rays are generally accepted to arise from the outer magnetosphere (cf. Cheng \& Zhang 1998; Takata et al. 2006, 2008).
The radio-quiet pulsars can be resulted from their different orientation of their radio beam and the observer's line-of-sight. 
Therefore, determining the emission 
geometry (e.g. magnetic inclination, viewing angle) is important for a better understanding of their properties. However, the lack of 
knowledge in the phase relationship between the $\gamma$-ray light curves and that in radio leads to an ambiguity in determining 
which peak in the $\gamma$-ray light curves is leading (see Trepl et al. 2010 for a more detailed discussion). Therefore, 
pulsation searches in the other wavelengths, in particular in X-ray, are needed. 

Only four of them have the X-ray pulsation detected so far (Halpern \& Holt 1992; Lin et al. 2010; Lin et al. 2013; 
Marelli et al. 2014).
%because of the limited exposures and the insufficient sensitivity for X-ray observatories with imaging capability.   
For those pulsars whose periodicities can only be detected in both X-ray and $\gamma$-ray bands are usually 
classified as Geminga-like pulsars with Geminga (i.e. PSR B0633+17) as the prototypical example of this class \citep{HH92,Bert92}.
X-ray pulsations of three other known Geminga-like pulsars, PSR~J0007+7302 \citep{Lin2010,Car2010}, 
PSR J2021+4026 \citep{Lin&Hui} and PSR~J1813-1246 (Marelli et al. 2014) were subsequently discovered.  
Except for PSR~J1813-1246 which has a very hard non-thermal dominant X-ray spectrum, 
the phase-averaged spectrum of other three Geminga-like pulsars have multicomponent which consists of a 
non-thermal contribution and at least one thermal component \citep{Car2004,Lin2010,Lin&Hui,Car2010}. 
The pulse profiles in X-ray and $\gamma$-ray are dissimilar with an obvious offset \citep{Car2010,Lin2010,Lin&Hui}.
This suggests the pulsed emission in these two regimes are originated from different regions. 
%Two tips or peaks associated with radiation originated from the particles accelerated region (e.g. outer gap of \citealt{TWC2011} or slot gap of \citealt{MH2003}) in the magnetosphere can be seen from the $\gamma$-ray folded light curve of a Geminga-like pulsar, but the X-ray one resembles a sinusoidal modulation that can be connected to a thermal surface emission.
%However, the complete physical properties of Geminga and two other Geminga-like pulsars inferred from the known periodicity and period derivative are different.
%PSR~J0007+7302 and PSR~J2021+4026 are much younger than Geminga, but they have stronger surface magnetic field \citep{Abdo2009a}.
%The spin-down power of PSR~J0007+7302 and PSR~J2021+4026 are also about one order higher than that of Geminga although they have the similar efficiency to convert spin-down power into pulsed $\gamma$-rays \citep{Abdo2008,Tre2010}.
%If we consider the characteristic age and the surface magnetic field of two Geminga-like pulsars, these two targets will be determined as Vela-like more properly.  
While the aforementioned three pulsars have similar X-ray/gamma-ray properties, we note that the derived paramaters (e.g. 
characteristic age, magnetic field and spin-down power) of PSR~J0007+7302 and PSR~J2021+4026 are rather different from Geminga. 

On the other hand, the X-ray counterpart of 3EG J1835+5928, RX J1836.2+5925, identified by \citet{MHEB2000} and \citet{MH2001} 
has been considered as the ``next Geminga" for a long time because of the lack of radio detection 
(a flux density of the radio observation at a frequency of 1.4 GHz $< 3\mu$Jy; \citealt{Abdo2010a}), 
the similar spectral behavior and $\gamma$-ray luminosity to Geminga \citep{HGMC2002}.
The $\gamma$-ray pulsation of this source was finally detected by \citet{Abdo2009a}, and it was classified as a 
middle-aged pulsar ($\sim 1800$~kyr) that is similar to Geminga.
The surface magnetic field ($\sim 5\times 10^{11}$~G) and the spin-down power 
($\sim 1.2\times 10^{34}$~erg s$^{-1}$) of \psr\ is about 1/3 to those of Geminga although its spin efficiency to 
convert $\gamma$-rays is about one order of magnitude higher \citep{Abdo2010a}.
However, the lack of the pulsation detected in other wavelengths does not allow us to compare the emission properties at different energies.  
Therefore, periodicity searches at other energies, particularly in X-ray band, are require to constrain its high energy emission 
properties and enable us to compare its nature with other Geminga-like pulsars. The X-ray pulsation have been long sought for this 
target \citep{HCG2007,Abdo2010a}. However, no positive detection has been reported yet.
%Although the previous {\it Chandra}/HRC (High Resolution Camera) observations of this target have accumulated exposures of $\sim 118$ ks, only $\sim 730-790$ counts were obtained for timing analysis because of the small effective area ($\sim 225$~cm$^{2}$).
%Even the long-term $\gamma$-ray ephemeris of this pulsar presents no significant timing noise and the periodic signal embedded in {\it Chandra} data can be expected with an extrapolation from the timing solution obtained by {\it Fermi}/LAT \citep{Abdo2010a}, no significant modulation was yielded from epoch-folding and an upper limit of 40\% on the pulsed fraction was concluded.  

Using the archival \emph{XMM-Newton} data of \psr, we have performed a detailed temporal and spectral analysis observations.
In this Letter, we report the results from this investigation. In particular, we report the discovery of the X-ray pulsations 
from \psr\ for the first time. 
 
\section{Observations and data analysis}

The latest {\it XMM-Newton} observations of \psr\ were on 2013 February 14 and 16 (hereafter 0214 and 0216) have a total exposure 
of $\sim$ 44 and 39 ks (Obs.~ID: 0693090101 and 0693090201; PI: Pavlov G.).   
During these observations, the satellite was pointed to RA=$18^{\rm h}36^{\rm m}13.75^{\rm s}$ Dec=$+59^{\circ}25^{'}30.3^{''}$ (J2000), which is the timing position of the $\gamma$-ray ephemeris determined by \citet{Abdo2010a}.
In both observations,  MOS1/2 CCDs were operated in full-window mode and PN CCD was operated in small-window mode with a 
temporal resolution of $\sim$5.7 ms which enables us to search for X-ray pulsations.  
We reprocessed all the raw data with the tasks \emph{emchain} and \emph{epchain} in XMM-Newton {\bf S}cience {\bf A}nalysis {\bf S}oftware (XMMSAS version 13.5.0). For a rigorous analysis, we have excluded those events next to the edges of CCDs and bad pixels, which may have incorrect 
energies.
The good events with ``PATTERN'' for MOS1/2 were selected in range of 0-12 to include single to quadruple pixel events, and those for the 
PN in the range 0-4 to include only single and double events. 
In our analysis, we only consider the events in the energy range of $0.2-12$~keV.
We also noted that all investigations have been contaminated by the X-ray background flare. 
After removing all events which are potentially contaminated, the effective exposures were 
found to be 41.1~ks, 41.3~ks and 41.4~ks for MOS1, MOS2, PN observed on 0214 and 32.3~ks, 32.2~ks 
and 32.8~ks for MOS1, MOS2, PN observed on 0216, respectively. We determined the position of \psr\ in each dataset with the 
XMMSAS task {\it edetect$\_$chain}. 
  
\subsection{Timing analysis}
For searching X-ray pulsations from \psr, we utilized the PN data obtained in both observations. 
Its nominal X-ray position in 0214 adn 0216 are  
R.A.=$18^h36^m13^s.68$, decl.=$+59^{\circ}$ $25'30''.72$ (J2000) (with the uncertainty $\sim 0''.5$) and  
R.A.=$18^h36^m13^s.92$, decl.=$+59^{\circ}25'29''.64$ (with the uncertainty $\sim 0''.7$).
Within the statisical uncertainties, they are consistent to the position determined by {\it Chandra}/ACIS (Advanced CCD Imaging Camera), 
{\it Chandra}/HRC (High Resolution Camera) \citep{HGMC2002,HCG2007} and the $\gamma$-ray timing position determined in \citet{Abdo2010a}. 
Since {\it Chandra}/HRC provides the most accurate positional determination, we adopted the position determined by this instrument,  
R.A.=$18^h36^m13^s.674$, decl.=$+59^{\circ}25'30''.15$ (J2000) for the barycentric correction. 
Events within a circular region of a 20$''$ radius centered at this position which corresponds to an encircle energy 
function of $\sim76\%$ were extracted. We have 992 counts on 0214 and 806 counts on 0216 available for the timing analysis.
We then corrected the photon arrival times to Barycentric Dynamical Time (TDB) with the aforementioned X-ray position and 
JPL Solar System ephemeris DE 405 using the XMMSAS task of {\it barycen}.  

%FFFFFFFFFFFFFFFFFFFFFFFFFFFFFFFFFFFFFFFFFF
\begin{figure}[t]
\centering
\includegraphics[width=8cm]{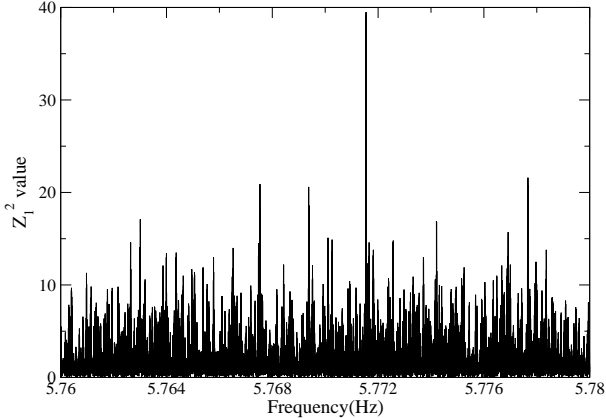}
\caption{{\footnotesize Detection of X-ray pulsation on PSR J1836+5925. 
The periodogram of Rayleigh-test in the frequency range of 5.76-5.78~Hz. 
The independent trial in the search corresponds to the Fourier resolution of this observation (i.e. $\sim 4.76\times 10^{-6}$~Hz).}}
\label{PDetection}
\end{figure}
%FFFFFFFFFFFFFFFFFFFFFFFFFFFFFFFFFFFFFFFFFF 
   
We noted that the $\gamma-$ray timing position of \psr\ in the available ephemerides has a small offset from 
our adopted X-ray position. For consistency, we constructed a new local ephemeris of this pulsar to cover the epoch 
of these recent {\it XMM-Newton} investigations (i.e. $\sim$MJD~56338). 
We used the {\it Fermi} Science Tools v9r27p1 package to perform the data reduction. 
We obtained the {\it Fermi} LAT data in the energy range of 0.1-300 GeV within a circular region of interest (ROI) with a $0.8^{\circ}$ radius from the decided X-ray position of \psr.
We used Pass 7 data and selected events in the ``Source" class (i.e.~event class 2).
Only a data span of $\sim1$~year, which brackets the epoch of two recent {\it XMM-Newton} observations, were 
considered to avoid accumulated effects due to the timing noise propagated from a long-term data.
We also excluded the events with zenith angles larger than 100$\degr$ to reduce the contamination by the Earth albedo $\gamma$-rays. 
For determining the pulse times of arrivals (TOAs), we built a template with the latest ephemeris reported by {\it Fermi} collaboration
\footnote{https://confluence.slac.stanford.edu/display/GLAMCOG/LAT+Gamma-ray+Pulsar+Timing+Models}
by means of Gaussian kernal density estimation (KDE).  
We have obtained 25 TOAs by cross-correlating the template with the unbinned geocentered data. Each event was assigned to a phase 
according to the KDE model. 
With the position fixed at that determined by \emph{Chandra}/HRC \citep{HCG2007}, we fitted the TOAs with \textsc{TEMPO2} \citep{HEM2006,EHM2006} 
to obtain a timing solution for \psr\ that includes a spin frequency ($f$), a spin-down rate ($\dot{f}$) and a second derivative of frequency 
($\ddot{f}$). The results are summarized in Table~\ref{ephemeris}.   

\begin{table}[t]
\caption{{\footnotesize Local ephemeris of \psr\ derived from LAT data which brackets the latest {\it XMM-Newton} observations on MJD~56337.45 and MJD~56339.44. The numbers in parentheses denote errors in the last digit.}}
\label{ephemeris}
\begin{tabular}{ll}
\hline\hline
{\footnotesize Pulsar name\dotfill} & {\footnotesize \psr} \\
{\footnotesize Valid MJD range\dotfill} & {\footnotesize 56147.8---56522.2} \\
{\footnotesize Right ascension, $\alpha$\dotfill} & {\footnotesize 18:36:13.674} \\
{\footnotesize Declination, $\delta$\dotfill} & {\footnotesize +59:25:30.15} \\
{\footnotesize Pulse frequency, $f$ (s$^{-1}$)\dotfill} & {\footnotesize 5.7715448470(3)} \\
{\footnotesize First derivative of pulse frequency, $\dot{f}$ (s$^{-2}$)\dotfill} & {\footnotesize $-5.002(3)\times10^{-14}$} \\
{\footnotesize second derivative of pulse frequency, $\ddot{f}$ (s$^{-3}$)\dotfill} & {\footnotesize $-1(1)\times10^{-23}$} \\
{\footnotesize Epoch of frequency determination (MJD)\dotfill} & {\footnotesize 56338} \\
{\footnotesize Solar system ephemeris model\dotfill} & {\footnotesize DE405} \\
{\footnotesize Time system \dotfill} & {\footnotesize TDB} \\
\hline
\end{tabular}
\end{table}

\begin{figure}[!t]
\centering
\includegraphics[width=8.5cm]{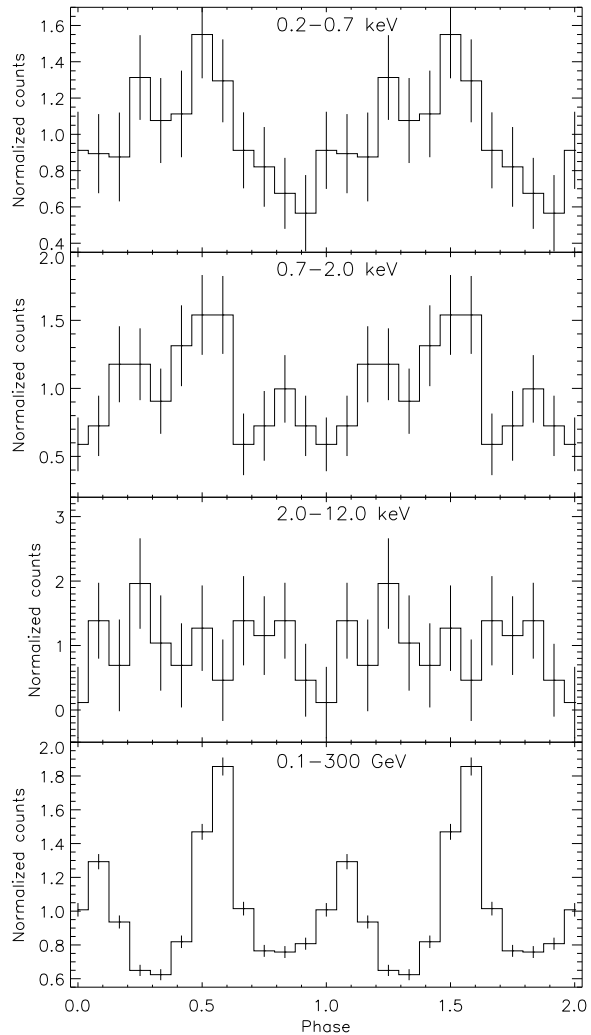}
\caption{{\footnotesize Folded light curves of \psr\ in different energy ranges. 
Each panel shows the pulse profile of 12 bins in the energy band specified in the legend. 
The epoch/phase zero of all the profiles were determined at MJD 56338 shown in Table~\ref{ephemeris}.}}
\label{PFXG}
\end{figure}

Using the $\gamma-$ray ephemeris which is contemporaneous with recent X-ray observations, we can compare temporal properties in both regimes. 
We directly folded up the PN data with the spin parameters in Table~\ref{ephemeris}. 
For the data in 0.2-12~keV, the random probability to yield the X-ray pulsation of 5.7715448470(3) Hz at epoch of 
MJD~56338 is $\sim 3.6\times 10^{-7}$ ($H$ test-statistic (TS) = 37.1) with a single harmonic (de Jager \& B\"{u}sching 2010).    
We have also computed the periodogram of Rayleigh-test in the frequency range 5.76-5.78 Hz which is shown in Figure~\ref{PDetection}. 
It clearly shows an X-ray periodic signal at 5.771545(5)~Hz with $Z^{2}_{1}=H~TS=39.5$. The quoted error of the frequency 
corresponds to the Fourier resolution of this observation \citep{Lea87}. This is consistent with the pulse frequency in $\gamma$-ray regime. 
To compare the pulse profiles between X-rays and $\gamma$-rays, we folded all the photons extracted from 
{\it XMM-Newton} observations and those obtained from {\it Fermi} observations within MJD~56150--56520 
in accordance with the $\gamma-$ray ephemeris Table~\ref{ephemeris}.
We also divided the obtained X-ray events into three different energy bands, which are similar to the studies  
of three other Geminga-like pulsars \citep{Car2004,Car2010,Lin&Hui}.
The phase-aligned folded light curves at different energy ranges are shown in Fig.~\ref{PFXG}. 
For the X-ray light curves, we have subtracted the background sampled from a nearby circular region with a radius of 20" centered 
at R.A.=$18^h36^m20^s.752$, decl.=$+59^{\circ}26'07''.92$ (J2000). 
The X-ray pulsation can be firmly detected in the soft X-ray band ($H$ TS=23.6 with a random probability of $\sim7.9\times10^{-5}$) 
and marginally found in the medium X-ray band ($H$ TS=11.6 with a random probability of $\sim 9.7\times10^{-3}$).
Nevertheless, the significance is too low for claiming the detection in the hard band 
($H$ TS=4.0 with a random probability of 0.2), which can be ascribed to the small photon statistic. 

%TTTTTTTTTTTTTTTTTTTTTTTTTTTTTTTTTTTTTTTTT
%TTTTTTTTTTTTTTTTTTTTTTTTTTTTTTTTTTTTTTTTT

\subsection{Spectral analysis}

Because our adopted data have the total exposures of more than 80~ks, this enables us to place a tighter constraint on the 
X-ray spectral properties of \psr\ than the previous investigations \citep{HGMC2002,Abdo2010a}.    
We extracted the spectra of \psr\ from a circle with a radius of $20''$ centered at its nominal position in each EPIC data. 
The background spectra were sampled from a nearby circular source-free region with a radius of $20''$ in individual dataset. 
We generated the response matrices and ancillary response files with the XMMSAS task {\it rmfgen} and {\it arfgen}.
Since the X-ray emission from \psr\ is relatively soft, we considered X-ray spectral fits in the energy range of 0.2--5~keV. 
We grouped each spectrum so as to have a minimum of 15 counts in each bin.  
%Because we have two observations investigated on different epoch, we examined the spectral behavior 
%of each one obtained from three cameras with XSPEC 12.8.1 in the beginning.
We have examined each dataset individually and found that the inferred parameters are consistent within the tolerance of the uncertainties.
In order to optimize the photon statistics, we fitted all the EPIC data obtained in these two observations simultaneously.     
The errors of the spectral parameters quoted in this Letter are in $1\sigma$ for 2 parameters of interest 
(i.e. $\Delta\chi^{2}=2.30$ above the minimum) for single component models and in $1\sigma$ for 4 parameters of interest 
(i.e. $\Delta\chi^{2}=4.72$ above the minimum) for multi-component models.

For the tested single component models (i.e. power-law and backbody), none of them can result in a statistically reasonable fit 
(with reduced $\chi^2>1.5$).
It may indicate that the X-ray emission of \psr\ comprises more than one spectral component. 
Therefore, we proceeded to fit the data with multicomponent spectral models. We have found that the spectrum can be described by
an absorbed blackbody plus power-law model. We firstly attempted to perform the fitting with all parameters to be free, which 
yields a column absorption of $n_{\rm H}<8.5\times10^{19}$~cm$^{-2}$, a photon index of $\Gamma=1.8\pm0.3$, a power-law model normalization 
of $6.0^{+1.3}_{-1.2}\times10^{-6}$~photons~keV$^{-1}$~cm$^{-2}$~s$^{-1}$ at 1 keV, a blackbody temperature of $kT=62^{+11}_{-10}$~eV with 
a emission radius of $R=1.32^{+2.37}_{-0.55}$~km at 800~pc. This results in a desirable goodness-of-fit ($\chi^{2}=90.46$ for 85 d.o.f.). 

\begin{figure}[t]
\centerline{\psfig{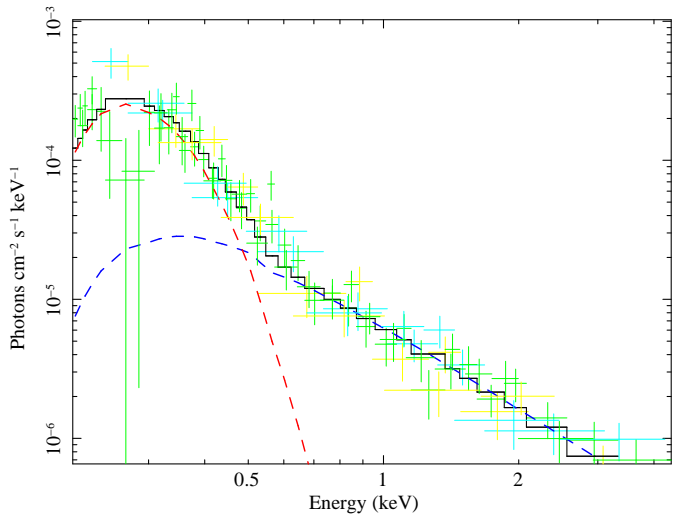}}
\caption[]{The phase-averaged X-ray spectra of \psr\ as observed by \emph{XMM-Newton} (MOS~1:cyan; MOS~2:yellow; PN:green) and
simultaneously fitted to an absorbed power-law plus blackbody model. The best-fit model for the thermal (red-dashed line)
and non-thermal (blue-dashed line) components are illustrated separately.}
\label{spec}
\end{figure}

As the column absorption cannot be properly constrained, we also fitted the data with 
$n_{\rm H}$ fixed at total Galactic HI column density of $4\times10^{20}$~cm$^{-2}$ in 
the direction of \psr\ \citep{Kalberla2005}. This yields the best-fit parameters of $\Gamma=2.1\pm0.3$, a power-law model normalization
of $(7.1\pm1.4)\times10^{-6}$~photons~keV$^{-1}$~cm$^{-2}$~s$^{-1}$ at 1 keV, $kT=45^{+10}_{-8}$~eV and a blackbody radius of 
$R=6.3^{+7.8}_{-3.4}$~km. The corresponding goodness-of-fit is $\chi^{2}=101.8$ for 86 d.o.f. which is statistically acceptable. 
The best-fit model and the observed spectra are shown in Figure~\ref{spec}. 

We have also considered the spectral fit with a dual blackbody model.
However, the goodness-of-fit yielded ($\chi^{2}=109.3$ for 85 d.o.f.) by this model is worse than the power-law plus blackbody model. 
Fixing the $n_{\rm H}$ at the Galactic HI value does not result in any improvement of the fitting. 
%We did not obtain any reasonable spectral fit when we fixed the column density at the known galactic value ($\chi^2=77.8$ for 47 d.o.f. and $\chi^2=57.7$ for 33 d.o.f.).
%The goodness of fit can be improved if the absorption was set to free. 
%Although the statistical results can only be marginally accepted, the temperature ($kT\sim 0.46$~keV) of the hotter component obtained from the fits listed in Table~\ref{spec} is comparable to that inferred from a heated polar cap \citep{CZ99}. 

\section{Discussion}

In this Letter, we report the discovery of X-ray pulsations from \psr. 
We found that the pulsed events both contribute in soft and medium bands (0.2--2~keV). On the other hand, 
no indication of pulsation in the hard band ($>$2~keV) has been found in our investigation.
The X-ray and $\gamma$-ray pulsation from \psr\ have significantly different behavior.
The pulsed $\gamma$-rays have a structure of relatively sharp double peaks which are presumably originated from 
the curvature radiation emitted from the pair creation region in the magnetosphere. 
The spin period of $P=173$~ms and the characteristic age of $\tau=1.8$~Myr suggest \psr\ as a mature pulsar.  
Assumming the emission is isotropic, 
the GeV gamma-ray efficiency of \psr\ is $\eta=(L_{\gamma}/\dot{E})\times 100\%=180\pm 1^{+200}_{-100}\%$ 
(second $Fermi$ pulsar catalog; \citealt{Abdo2013}), where $\dot{E}$ is the spin down power. 
The first uncertainty of $\eta$ comes from the statistical uncertainties in the spectral fit and the second is due to
the distance uncertainty \citep{Abdo2013}.
This high efficiency of the GeV emission is similar to $\eta=97.4\pm0.3^{+300}_{-50}\%$ of typical mature pulsar --- Geminga, 
if one assumes a 4$\pi$ solid angle. Considering the possible
overestimated distances and/or the beaming factor, the aforementioned conversion efficiencies can be overestimated. Taking the 
more realistic parameters into account, this might result in an efficiency $\eta<100$\%. 

In order to compare with Geminga, therefore, we discuss the origin of the X-ray emission of \psr.
In the soft band (see Fig~\ref{PFXG}), the light curve is essentially sinusoidal which suggests the X-ray pulsation is mostly originated 
from the modulation of the heated polar cap. This is consistent with the presence of a blackbody component inferred from the spectral fit. 

To account for the power-law component, we discuss it in the context of the outergap model \citep{CZ98}. 
According to this model, the non-thermal component of a canonical pulsar is synchrotron emission from the relativistic
$e^{-}/e^{+}$ in the outer magnetospheric gap \citep{TSHC2006,TCS2008} with a week absorption.
The photon index is expected to be in the range of 1.5--2.0 which is consistent with the best-fit value. 
For the light curve in 0.7-2~keV, its profile is apparently different from that in 0.2-0.7~keV which might suggests a different origin.
%In particular, there is indication for the presence of mulitple peaks which apparently align with the $\gamma$-ray peaks (see Fig~\ref{PFXG}) 
%which suggests the pulsed X-rays in the medium band are dominated by the synchrotron radiation of the secondary electron and positron pairs
%produced in the pulsar magnetosphere \citep{ZC97}. 
In examining the relative 
flux contributions by the thermal and non-thermal component, we found that while the non-thermal flux takes $\sim15\%$ in 0.2-0.7~keV its 
contribution is $>99\%$ in 0.7-2~keV (see also Fig.~\ref{spec}). 
This suggests that X-ray pulsations seen in 0.2-0.7~keV and 0.7-2~keV might have different origin. 
However, the small photon statistics in the medium band precludes any firm conclusion. 
To better characterize the X-rays from \psr\ in different energy bands, a deeper observation is certainly required. With a larger number
of photons detected, we can also perform a phase-resolved spectroscopy to investigate how do the spectral properties vary across
the rotational phase which will provide important inputs for a physical modeling of the pulsar. 
%For comparison, the observed X-ray emission of Geminga is described by two blackbody component plus power-law component, 
%and the power-law component dominates blackbody components at energies $\gtrsim1$~keV. 

%FFFFFFFFFFFFFFFFFFFFFFFFFFFFFFFFFFFFFFFFFF
%\begin{figure}[t]
%\centering
%\includegraphics[width=6.0cm,angle=-90]{po+bb.eps}
%\caption{{\footnotesize Confidence contours for the dual-components fit to the {\it XMM-Newton} spectrum of \psr.  
%The column density in the spectral model has been fixed at two different values for an obvious comparison, and one of them was set as the galactic absorption obtained from \citet{Kalberla2005}.
%Confidence levels in the plot are for two interesting parameters.}}
%\label{contour}
%\end{figure}
%FFFFFFFFFFFFFFFFFFFFFFFFFFFFFFFFFFFFFFFFFF 

\begin{acknowledgements} 
We thank Mr. Jason H. K. Wu of Max-Planck-Institut f$\ddot{u}$r Radioastronomie for providing useful comments on the temporal analysis 
techniques.
This work made use of data supplied by the High Energy Astrophysics Science Archive Research Center (HEASARC). 
This project is partially supported by the Ministry of Science and Technology of Taiwan through grant NSC~101-2112-M-039-001-MY3.
C.~Y~.H. is supported by the research fund of Chungnam National University in 2014.
J.~T. is supported by a GRF grant of Hong-Kong Government under HKU700911P.
C.~P.~H. and Y.~C. are supported by the Ministry of Science and Technology of Taiwan through the grant NSC 102-2112-M-008-020-MY3.
A.~K.~H.~K. is supported by the Ministry of Science and Technology of Taiwan through grants NSC~100-2628-M-007-002-MY3 and NSC~100-2923-M-007-001-MY3.  
D.~C.~C.~Y. gets the financial support from the Ministry of Science and Technology of Taiwan through the grant NSC~102-2115-M-030-003 and the other FJU project of A0502004.
\end{acknowledgements}


\begin{thebibliography}{21}
\expandafter\ifx\csname natexlab\endcsname\relax\def\natexlab#1{#1}\fi

\bibitem[{{Abdo} {et~al.}(2009){Abdo}, {Ackermann}, {Ajello}, {Anderson},
  {Atwood}, {Axelsson}, {Baldini}, {Ballet}, {Barbiellini}, {Baring},
  {Bastieri}, {Baughman}, {Bechtol}, {Bellazzini}, {Berenji}, {Bignami},
  {Blandford}, {Bloom}, {Bonamente}, {Borgland}, {Bregeon}, {Brez}, {Brigida},
  {Bruel}, {Burnett}, {Caliandro}, {Cameron}, {Caraveo}, {Casandjian},
  {Cecchi}, {{\c C}elik}, {Chekhtman}, {Cheung}, {Chiang}, {Ciprini}, {Claus},
  {Cohen-Tanugi}, {Conrad}, {Cutini}, {Dermer}, {de Angelis}, {de Luca}, {de
  Palma}, {Digel}, {Dormody}, {do Couto e Silva}, {Drell}, {Dubois}, {Dumora},
  {Farnier}, {Favuzzi}, {Fegan}, {Fukazawa}, {Funk}, {Fusco}, {Gargano},
  {Gasparrini}, {Gehrels}, {Germani}, {Giebels}, {Giglietto}, {Giommi},
  {Giordano}, {Glanzman}, {Godfrey}, {Grenier}, {Grondin}, {Grove},
  {Guillemot}, {Guiriec}, {Gwon}, {Hanabata}, {Harding}, {Hayashida}, {Hays},
  {Hughes}, {J{\'o}hannesson}, {Johnson}, {Johnson}, {Johnson}, {Kamae},
  {Katagiri}, {Kataoka}, {Kawai}, {Kerr}, {Kn{\"o}dlseder}, {Kocian}, {Kuss},
  {Lande}, {Latronico}, {Lemoine-Goumard}, {Longo}, {Loparco}, {Lott},
  {Lovellette}, {Lubrano}, {Madejski}, {Makeev}, {Marelli}, {Mazziotta},
  {McConville}, {McEnery}, {Meurer}, {Michelson}, {Mitthumsiri}, {Mizuno},
  {Monte}, {Monzani}, {Morselli}, {Moskalenko}, {Murgia}, {Nolan}, {Norris},
  {Nuss}, {Ohsugi}, {Omodei}, {Orlando}, {Ormes}, {Paneque}, {Parent},
  {Pelassa}, {Pepe}, {Pesce-Rollins}, {Pierbattista}, {Piron}, {Porter},
  {Primack}, {Rain{\`o}}, {Rando}, {Ray}, {Razzano}, {Rea}, {Reimer}, {Reimer},
  {Reposeur}, {Ritz}, {Rochester}, {Rodriguez}, {Romani}, {Ryde},
  {Sadrozinski}, {Sanchez}, {Sander}, {Parkinson}, {Scargle}, {Sgr{\`o}},
  {Siskind}, {Smith}, {Smith}, {Spandre}, {Spinelli}, {Starck}, {Strickman},
  {Suson}, {Tajima}, {Takahashi}, {Takahashi}, {Tanaka}, {Thayer}, {Thompson},
  {Tibaldo}, {Tibolla}, {Torres}, {Tosti}, {Tramacere}, {Uchiyama}, {Usher},
  {Van Etten}, {Vasileiou}, {Vilchez}, {Vitale}, {Waite}, {Wang}, {Watters},
  {Winer}, {Wolff}, {Wood}, {Ylinen}, \& {Ziegler}}]{Abdo2009a}
{Abdo}, A.~A., {et~al.} 2009, Science, 325, 840

\bibitem[{{Abdo} {et~al.}(2010){Abdo}, {Ackermann}, {Ajello}, {Atwood},
  {Baldini}, {Ballet}, {Barbiellini}, {Baring}, {Bastieri}, {Bechtol},
  {Belfiore}, {Bellazzini}, {Berenji}, {Blandford}, {Bloom}, {Bonamente},
  {Borgland}, {Bregeon}, {Brez}, {Brigida}, {Bruel}, {Burnett}, {Buson},
  {Caliandro}, {Cameron}, {Camilo}, {Caraveo}, {Carrigan}, {Casandjian},
  {Cecchi}, {{\c C}elik}, {Charles}, {Chekhtman}, {Cheung}, {Chiang},
  {Ciprini}, {Claus}, {Cohen-Tanugi}, {Conrad}, {de Angelis}, {de Luca}, {de
  Palma}, {Digel}, {Dormody}, {Silva}, {Drell}, {Dubois}, {Dumora}, {Edmonds},
  {Farnier}, {Favuzzi}, {Fegan}, {Focke}, {Fortin}, {Frailis}, {Fukazawa},
  {Funk}, {Fusco}, {Gargano}, {Gasparrini}, {Gehrels}, {Germani}, {Giavitto},
  {Giglietto}, {Giordano}, {Glanzman}, {Godfrey}, {Grenier}, {Grondin},
  {Grove}, {Guillemot}, {Guiriec}, {Gwon}, {Hadasch}, {Harding}, {Hays},
  {Horan}, {Hughes}, {Jackson}, {J{\'o}hannesson}, {Johnson}, {Johnson},
  {Johnson}, {Johnson}, {Kamae}, {Kanai}, {Katagiri}, {Kataoka}, {Kawai},
  {Kerr}, {Kn{\"o}dlseder}, {Kuss}, {Lande}, {Latronico}, {Lemoine-Goumard},
  {Longo}, {Loparco}, {Lott}, {Lovellette}, {Lubrano}, {Madejski}, {Makeev},
  {Marelli}, {Mazziotta}, {McEnery}, {Meurer}, {Michelson}, {Mitthumsiri},
  {Mizuno}, {Moiseev}, {Monte}, {Monzani}, {Morselli}, {Moskalenko}, {Murgia},
  {Nolan}, {Norris}, {Nuss}, {Ohsugi}, {Omodei}, {Orlando}, {Ormes}, {Paneque},
  {Panetta}, {Parent}, {Pelassa}, {Pepe}, {Pesce-Rollins}, {Pierbattista},
  {Piron}, {Porter}, {Rain{\`o}}, {Rando}, {Ransom}, {Ray}, {Razzano}, {Rea},
  {Reimer}, {Reimer}, {Reposeur}, {Rochester}, {Rodriguez}, {Romani}, {Roth},
  {Ryde}, {Sadrozinski}, {Sander}, {Saz Parkinson}, {Scargle}, {Sgr{\`o}},
  {Siskind}, {Smith}, {Smith}, {Spandre}, {Spinelli}, {Strickman}, {Suson},
  {Takahashi}, {Tanaka}, {Thayer}, {Thayer}, {Thompson}, {Thorsett}, {Tibaldo},
  {Tibolla}, {Torres}, {Tosti}, {Tramacere}, {Usher}, {Van Etten}, {Vasileiou},
  {Venter}, {Vilchez}, {Vitale}, {Waite}, {Wang}, {Watters}, {Winer}, {Wolff},
  {Wood}, {Ylinen}, \& {Ziegler}}]{Abdo2010a}
---. 2010, \apj, 712, 1209

\bibitem[{{Abdo} {et~al.}(2013){Abdo}, {Ajello}, {Allafort}, {Baldini},
  {Ballet}, {Barbiellini}, {Baring}, {Bastieri}, {Belfiore}, {Bellazzini}, \&
  et~al.}]{Abdo2013}
---. 2013, \apjs, 208, 17

\bibitem[{{Bertsch} {et~al.}(1992){Bertsch}, {Brazier}, {Fichtel}, {Hartman},
  {Hunter}, {Kanbach}, {Kniffen}, {Kwok}, {Lin}, \& {Mattox}}]{Bert92}
{Bertsch}, D.~L., {et~al.} 1992, \nat, 357, 306

\bibitem[{{Caraveo} {et~al.}(2010){Caraveo}, {De Luca}, {Marelli}, {Bignami},
  {Ray}, {Saz Parkinson}, \& {Kanbach}}]{Car2010}
{Caraveo}, P.~A., {De Luca}, A., {Marelli}, M., {Bignami}, G.~F., {Ray}, P.~S.,
  {Saz Parkinson}, P.~M., \& {Kanbach}, G. 2010, \apjl, 725, L6

\bibitem[{{Caraveo} {et~al.}(2004){Caraveo}, {De Luca}, {Mereghetti},
  {Pellizzoni}, \& {Bignami}}]{Car2004}
{Caraveo}, P.~A., {De Luca}, A., {Mereghetti}, S., {Pellizzoni}, A., \&
  {Bignami}, G.~F. 2004, Science, 305, 376

\bibitem[{{Cheng} \& {Zhang}(1998)}]{CZ98}
{Cheng}, K.~S., \& {Zhang}, L. 1998, \apj, 498, 327

\bibitem[{{Edwards} {et~al.}(2006){Edwards}, {Hobbs}, \&
  {Manchester}}]{EHM2006}
{Edwards}, R.~T., {Hobbs}, G.~B., \& {Manchester}, R.~N. 2006, \mnras, 372,
  1549

\bibitem[{{Halpern} {et~al.}(2007){Halpern}, {Camilo}, \& {Gotthelf}}]{HCG2007}
{Halpern}, J.~P., {Camilo}, F., \& {Gotthelf}, E.~V. 2007, \apj, 668, 1154

\bibitem[{{Halpern} {et~al.}(2002){Halpern}, {Gotthelf}, {Mirabal}, \&
  {Camilo}}]{HGMC2002}
{Halpern}, J.~P., {Gotthelf}, E.~V., {Mirabal}, N., \& {Camilo}, F. 2002,
  \apjl, 573, L41

\bibitem[{{Halpern} \& {Holt}(1992)}]{HH92}
{Halpern}, J.~P., \& {Holt}, S.~S. 1992, \nat, 357, 222

\bibitem[{{Hobbs} {et~al.}(2006){Hobbs}, {Edwards}, \& {Manchester}}]{HEM2006}
{Hobbs}, G.~B., {Edwards}, R.~T., \& {Manchester}, R.~N. 2006, \mnras, 369, 655

\bibitem[{{Kalberla} {et~al.}(2005){Kalberla}, {Burton}, {Hartmann}, {Arnal},
  {Bajaja}, {Morras}, \& {P{\"o}ppel}}]{Kalberla2005}
{Kalberla}, P.~M.~W., {Burton}, W.~B., {Hartmann}, D., {Arnal}, E.~M.,
  {Bajaja}, E., {Morras}, R., \& {P{\"o}ppel}, W.~G.~L. 2005, \aap, 440, 775

\bibitem[{{Kerr}(2011)}]{Kerr2011}
{Kerr}, M. 2011, \apj, 732, 38

\bibitem[{{Leahy}(1987)}]{Lea87}
{Leahy}, D.~A. 1987, \aap, 180, 275

\bibitem[{{Lin} {et~al.}(2010){Lin}, {Huang}, {Takata}, {Hwang}, {Kong}, \&
  {Hui}}]{Lin2010}
{Lin}, L.~C.~C., {Huang}, R.~H.~H., {Takata}, J., {Hwang}, C.~Y., {Kong},
  A.~K.~H., \& {Hui}, C.~Y. 2010, \apjl, 725, L1

\bibitem[{{Lin} {et~al.}(2013){Lin}, {Hui}, {Hu}, {Wu}, {Huang}, {Trepl},
  {Takata}, {Seo}, {Wang}, {Chou}, \& {Cheng}}]{Lin&Hui}
{Lin}, L.~C.~C., {et~al.} 2013, \apjl, 770, L9

\bibitem[{{Marelli} {et~al.}(2014){Marelli}, {Harding}, {Pizzocaro}, {De Luca},
  {Wood}, {Caraveo}, {Salvetti}, {Saz Parkinson}, \& {Acero}}]{Marelli2014}
{Marelli}, M., {et~al.} 2014, ArXiv e-prints

\bibitem[{{Mirabal} \& {Halpern}(2001)}]{MH2001}
{Mirabal}, N., \& {Halpern}, J.~P. 2001, \apjl, 547, L137

\bibitem[{{Mirabal} {et~al.}(2000){Mirabal}, {Halpern}, {Eracleous}, \&
  {Becker}}]{MHEB2000}
{Mirabal}, N., {Halpern}, J.~P., {Eracleous}, M., \& {Becker}, R.~H. 2000,
  \apj, 541, 180

\bibitem[{{Takata} {et~al.}(2008){Takata}, {Chang}, \& {Shibata}}]{TCS2008}
{Takata}, J., {Chang}, H., \& {Shibata}, S. 2008, \mnras, 386, 748

\bibitem[{{Takata} {et~al.}(2006){Takata}, {Shibata}, {Hirotani}, \&
  {Chang}}]{TSHC2006}
{Takata}, J., {Shibata}, S., {Hirotani}, K., \& {Chang}, H.-K. 2006, \mnras,
  366, 1310

\bibitem[{{Trepl} {et~al.}(1987)}]{Trepl10}
{Trepl}, L. et al. 2010, \mnras, 405, 1339

\bibitem[{{Zhang} \& {Cheng}(1997)}]{ZC97}
{Zhang}, L., \& {Cheng}, K.~S. 1997, \apj, 487, 370

\end{thebibliography}
\end{document}